\documentclass[aps,floats,prd]{revtex4}
\usepackage{amsmath}
\usepackage{amsfonts,amsmath,mathrsfs}
\usepackage{epsfig}
\usepackage{latexsym}
\usepackage{bbm}
\usepackage{amssymb}
\usepackage{graphicx}
\usepackage{dcolumn}
\usepackage{bm}

\begin{document}

\title{An Emergent World of Gauge Force and Partons}
\author{Yao Ma}
\author{Zheng-Yu Weng}
\address{Institute for Advanced Study, Tsinghua University\\
Beijing, 100084, China}


\begin{abstract} 
We illustrate how a completely new world of gauge force emerges from a conventional condensed matter system in a rigorous way. A characteristic energy scale (Mott gap) separates such an exotic universe from the ordinary one that we condensed matter physicists are more familiar with at higher energies. The governing physical law is no longer about individual electrons but concerns fractionalized particles, i.e., partons, as the new collective modes resulted from strong correlation among the electrons. Novel phenomena in this low-energy universe are clearly distinguished from Landau's Fermi liquid described by the perturbative quantum many-body theory.

\end{abstract}

\date{}
\maketitle

\section{Introduction}


In 1974, C. N. Yang\cite{Yang1974} pointed out that a fundamental gauge force is completely and intrinsically characterized by a non-integrable phase factor that an elementary particle of charge $q$ picks up when it moves from A to B points:
\begin{equation}
{P}\exp \left( i\frac{q}{\hbar c}\int^B_A A_{\mu}dx^{\mu}\right) .
\label{nonit}
\end{equation}
Here $A_{\mu}$ is the gauge potential and $P$ stands for path ordering. Today it has been fully established that the fundamental forces of weak, electromagnetic, and strong interactions between the elementary particles are all dictated by (\ref{nonit}) as gauge forces, which is a great triumph of physics in the twentieth century. 

The world of condensed matter physics is within the realm of a particular fundamental force, namely, the electromagnetic force. In the presence of millions of other electrons and ionic atoms, an electron moving in a solid is no longer a free `bare' electron in the vacuum.  Nevertheless, the electron still remains `free' as a Bloch electron in a periodic array of the ionic atoms as if the latter is `transparent', only with its energy spectrum being altered into the so-called energy bands. If the lattice vibration and the screened Coulomb repulsion from other electrons can be treated perturbatively, the Bloch electron world in the condensed matter, known as the Landau's Fermi liquid state, is well described by the quantum many-body theory which was a great achievement in '50.\cite{AGD1963,Pines1965}

But Nature is more complex and richer. In some real materials, the band mass of a Bloch electron can become much heavier or its bandwidth is much reduced as compared to the strength of the local Coulomb interaction between the electrons.  This is precisely the proposal made by P. W. Anderson\cite{pwa} in 1987 in explaining high-$T_c$ superconductivity discovered in the cuprate compounds. The proposed model for the cuprates is extremely simple. That is, the single-band Bloch electrons in the copper-oxide layers of the cuprate will experience a strong on-site Coulomb repulsion $U$ as compared to a relatively narrow bandwidth. This Hubbard-type model is a great simplification of a more realistic description of the copper-oxides, justified at reasonable energy and length scales from quantum chemistry. An entirely new physics is conjectured\cite{pwa1} to arise in the infrared regime. But the problem is that the well-established perturbation-based quantum many-body theory\cite{AGD1963,Pines1965} is expected to completely fail in this strong correlation limit. The challenge to theorists is how to identify a right mechanism, and at the same time, to find an appropriate mathematical description for such a perturbation inoperative regime, in order to accommodate a great amount of anomalous properties observed in cuprate superconductors.\cite{pwa1,LNW2006} 

The story that we will present below is that for such an innocent-looking large-$U$ Hubbard model, its infrared novelty can be fully captured by an exotic emergent gauge force in the same fashion that the elementary particles experience the fundamental forces in Nature. That is, novel physical properties observed in experiments may be essentially attributed to some relevant low-energy collective degrees of freedom with an intrinsic gauge structure. In this sense, the emergent gauge force in the cuprates is for real, which is precise and robust when the length scale is larger than the lattice constant of the copper-oxide unit cells and the energy is much less than a characteristic scale caused by the Hubbard $U$ for two electrons staying at the same site (i.e., the so-called Mott gap $> 2$$eV$\cite{Wang2012}).

Here it is crucial to distinguish such an intrinsic gauge force from a fictitious gauge degree of freedom arising from a formal decomposition of an electron operator mathematically, e.g., in the so-called slave-particle formulations.\cite{LNW2006} The latter represents a counter-effect against breaking up the electron degrees of freedom and disappears if one stays with the electronic description. In other words, the spinless holon of charge $+e$ and neutral spinon of spin-$1/2$ that appear in the fractionalizationof an electron carry opposite gauge charges such that the electron as a composite object is still gauge-neutral. In particular, these latter gauge fluctuations will be suppressed or `Higgsed' when the holons are condensed and/or spinons are in singlet pairing\cite{LNW2006} [i.e., the so-called resonating-valence-bond (RVB) state\cite{pwa}]. 

By contrast, the intrinsic gauge force discussed in the following will exist generally as a precise property\cite{Weng1996,Weng1997,WWZ2008} hidden in the $t$-$J$ model (the large-$U$ Hubbard model), which will dictate how the physical degrees of freedom are fractionalized mathematically. In particular, the electrons can always feel the effect of such a gauge force, even if the subsystems (partons) of fractionalization have experienced off-diagonal-long-range-orders (ODLROs), like holon condensation and/or spinon RVB pairing. As such, it is fundamentally different from the aforementioned gauge fluctuations\cite{LNW2006} associated with the slave-particle fractionalization, which can be effectively eliminated by the ODLROs of the partons. This novel gauge force plays a critical role\cite{MW2010} in driving superconducting phase transition of the electronic system. Consequently the system exhibits non-BCS superconducting and non-Fermi-liquid normal-state behavior because of such emergent gauge fields.\cite{Weng2007,Zaanen2009,Weng2011r} 

\section{Non-integrable phase factor: Emergent gauge force} 

\begin{figure}[t]
\begin{center}
\includegraphics[width=0.6\textwidth]{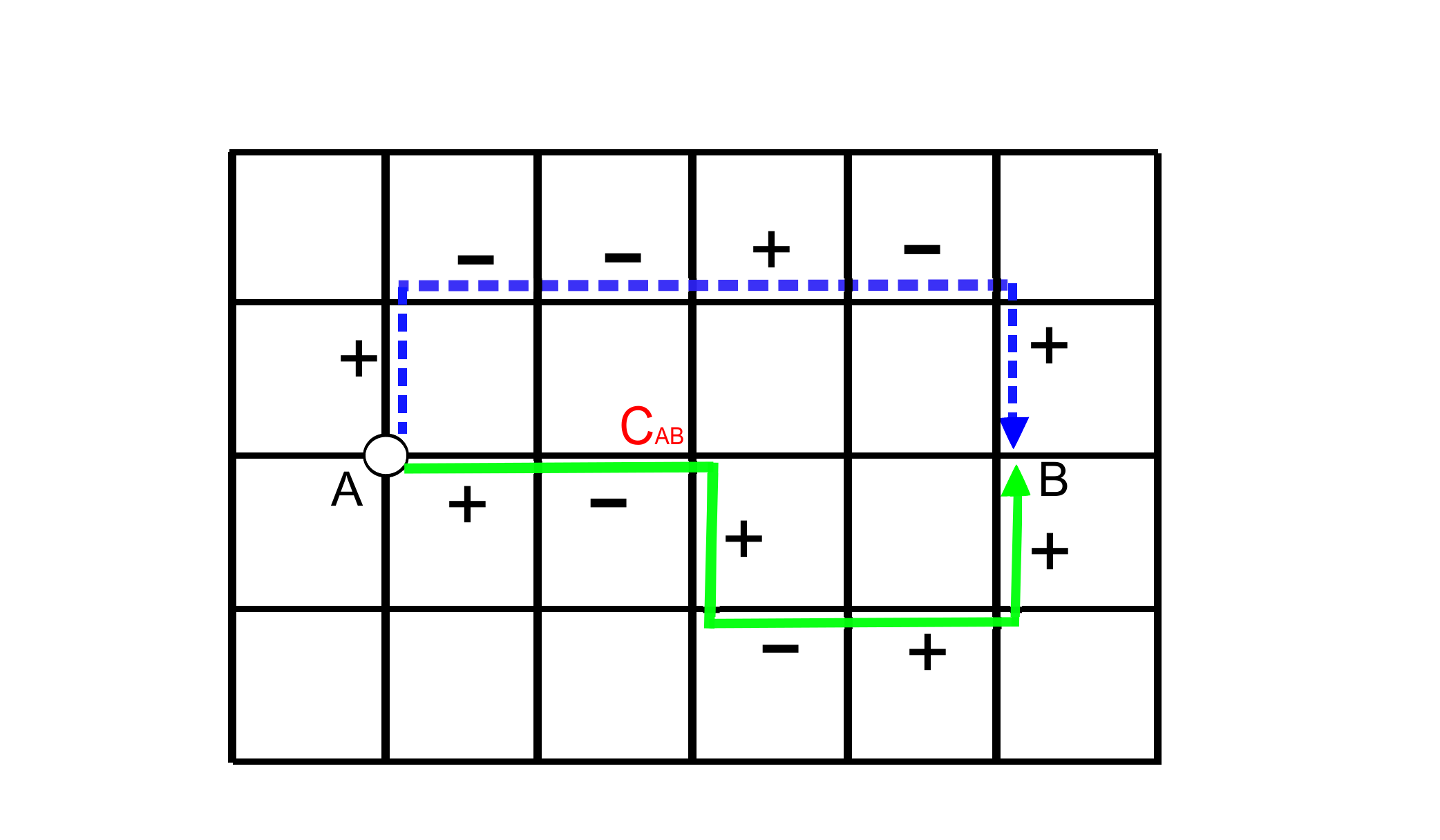}
\end{center}
\caption{Non-integrable phase factor as a product of signs, defined in (\ref{ps}), is equivalent to an emergent gauge force. It is picked up by a hole moving from A to B along a given path $C_{AB}$, and its novel quantum interference effect will be responsible for the essential physics of a doped Mott insulator. }
\label{psfig}
\end{figure}

Let us start by outlining an exact theorem for the large-$U$ Hubbard model, i.e., the $t$-$J$ model on a bipartite lattice, which holds true for arbitrary dimensions, electron concentration, and temperature.\cite{Weng1996,Weng1997,WWZ2008} It states that when a charge carrier moves from A to B, it will always pick up a phase factor as a product of signs:
\begin{equation}
\tau_{C_{AB}}\equiv (+1)\times (-1)\times (-1)\times \cdot\cdot\cdot 
\label{ps}
\end{equation}
As illustrated in Fig. 1, here $\tau_{C_{AB}}$ is path-dependent, and thus is a typical \emph{non-integrable} phase factor; Each $\pm $ sign faithfully records the doublet ($\uparrow$ or $\downarrow$) spin backflow at each step of hopping to the nearest-neighbor site (a neutral spin must be simultaneously exchanged positions with the charge according to the $t$-$J$ model); Such a non-integrable phase factor is known as the phase string effect,\cite{Weng1996,Weng1997} which appears in the single-particle propagator, total energy expression, and the partition function of the system. For example, the partition function of the $t$-$J$ model can be generally expressed by\cite{WWZ2008}
\begin{equation}
Z=\sum_{\{C\}}\tau_{C }\tau^{hh}_C {W}[\{C\}]
\label{Z}
\end{equation}
where $\{C\}$ represents the closed loops of the whole charge carriers and spins;  The positive weight ${W}[\{C\}]>0$
depends on all the detailed coupling constants as well as the dimensionality, temperature, and electron concentration; Finally an additional sign factor
\begin{equation}
\tau^{hh}_{C}\equiv (-1)\times (-1)\times  (-1)\times \cdot\cdot\cdot
\label{hh}
\end{equation}
represents conventional Fermionic statistical signs resulted from exchanges between the doped charge carriers.\cite{WWZ2008} Near half-filling, such statistical signs are much sparser than the phase string signs (\ref{ps}) picked up by the charge carriers over the set of closed loops $\{C\}$.

The important implication is as follows. The phase string factor (\ref{ps}) as a non-integrable phase factor proves that the long-wavelength physics of the large-$U$ Hubbard model or the $t$-$J$ model is completely and intrinsically dictated by a gauge force, just like the general definition (\ref{nonit}) for the fundamental forces of Nature. Such a gauge force is emergent by nature because it disappears when the Hubbard $U$ is reduced to much smaller than the bandwidth, where the perturbative quantum many-body theory works and the system is well described by the Landau's Fermi liquid theory. In this weak coupling limit, the solid state is conventional with the electrons behaving as normal fermions, and may be continuously connected to the Fermi gas limit of the Bloch electrons. But the strong coupling limit, as characterized by the opening up of a Mott gap, signals\cite{note1} the failure of the quantum many-body theory and the Landau's paradigm. Here a new paradigm in the condensed matter emerges.\cite{Zaanen2009,Weng2011r} It is a universe in which individual electrons themselves are no longer elementary but `emergent', while the basic constituent degrees of freedom are collective in nature, including both charge and spin, and the fundamental force becomes a long-range gauge force.\cite{KQW2005,Ye2011}

The nature of such an emergent gauge force can be understood by noting that the non-integrable sign factor (\ref{ps}) leads to exotic quantum interference effect for a charge particle traversing a quantum spin background from A to B. Namely, the spin degrees of freedom will strongly influence the charge degree of freedom nonlocally via (\ref{ps}). A very illustrative example will be given below. \emph{ Vice versa}, the spin degrees of freedom will be strongly influenced by (\ref{ps}) as well in the presence of a finite density of the charge carriers in order to optimize the \emph{total} energy of the system. In other words, the gauge force as introduced by (\ref{ps}) will mediate a long-range \emph{ mutual} entanglement between the charge and spin degrees of freedom, leading to all the novelty in the different concentration, temperature, etc., regimes of a complex phase diagram. Such an emergent gauge structure is described mathematically by a mutual Chern-Simons gauge fields.\cite{KQW2005,Ye2011}

The non-integrable phase factor appearing in (\ref{Z}) indicates that the fermion signs of the electrons are completely altered by the strong interaction. In fact, the fermion signs totally disappear at half-filling limit, with $\tau_{C }$=$\tau^{hh}_C =1$, which is the simplest but most striking example of how the strong interaction can completely change the electron statistics. Sparse signs pop up with the introduction of doped holes via  $\tau_{C }$ and $\tau^{hh}_C$, which are much reduced at low doping as compared to the original fermionic signs associated with the underlying electrons. 

The simplest nontrivial case is doping one hole into the half-filling Mott insulator. Here one has $\tau^{hh}_C =1$ and the phase string factor $\tau_{C }$ plays singularly the most important role. Generally, one expects that when the spatial distance between the given A and B points in Fig. 1 is sufficiently long, the non-integrable phase factor $\tau_{C _{AB}}$ in (\ref{ps}) can result in a severe destructive quantum interference from different paths with $\tau_{C _{AB}}$ differing by $\pm 1$. As an analogy to Anderson localization,\cite{Anderson1958} the single hole doped in such a Mott insulator will be self-localized\cite{Weng2000} (see below). This is probably one of the most striking consequences of the non-integrable phase factor identified for the $t$-$J$ model: In sharp contrast to Anderson localization, the present system is translation invariant without the presence of disorder and the entire novelty comes from the emergent gauge force as the strong correlation effect.

Summing up all the paths with singular $\tau_{C _{AB}}$ at long distance is a difficult analytic task. However, to reveal the novel quantum interference effect of the non-integrable phase factor (\ref{ps}),   one may consider a ladder system with one direction of the sample very long, but finite along the other direction, instead of a large two-dimensional square lattice. As a matter of fact, even for a two-leg ladder, i.e.,  two one-dimensional chains coupled together, there are more than one path to connect any A and B lattice sites, and consequently, $\tau_{C _{AB}}$ becomes non-integrable phase factor to play the crucial role of quantum interference so long as the ladder is long enough along the chain direction. Fortunately for the ladder systems, the density matrix renormalization group (DMRG) method provides an accurate and powerful numerical approach to this problem.\cite{White1992}  A systematic DMRG study on the one-hole problem in large-scale $t$-$J$ ladder systems, with the leg number ranging from 1 to 5, has been carried out recently\cite{Zhu2012}. The results have unequivocally demonstrated that the single hole is indeed self-localized once the leg number is larger than one. By contrast, once the non-integrable phase factor $\tau_{C _{AB}}$ is turned off by slightly modifying\cite{Zhu2012} the $t$-$J$ Hamiltonian, the self-localization disappears and the doped hole recovers a well-defined Bloch quasiparticle behavior. 

\section{Ground state wave function that precisely incorporates the non-integrable phase factor}

So far no approximation has been made about the model and one has already known deeply about its novel nature by identifying the non-integrable phase factor (\ref{ps}). In essence, the non-integrable phase factor of $\tau_{C }$ and $\tau^{hh}_C$ in (\ref{Z}) is topological rather than geometrical, in the sense that it is completely specified by the numbers of hole-spin exchanges, $N_h^{\downarrow}(\{C\})$, and hole-hole exchanges, $N^h_h(\{C\})$, for a given set of closed paths $\{C\}$. Therefore, such a sign structure resembles the \emph{altered} statistical signs. In other words, the familiar fermion signs in a weakly interacting electron system, which is responsible for the Fermi liquid behavior, is now replaced by new statistical signs encoded by $\tau_{C }$ and $\tau^{hh}_C$.

One may introduce a unitary transformation $\hat{K}$ to reexpress the ground state of the $t$-$J$ model by
\begin{equation}
|\Psi _{\mathrm{G}}\rangle =\hat{K} |{\Phi} _{\mathrm{G}}\rangle ~ .
\label{unitary}
\end{equation}
Here the so-called mutual duality transformation $\hat{K}$ will properly incorporate the precise phase string (\ref{ps}), which is statatical in nature as discussed above. On the other hand, $\tau^{hh}_C$ in (\ref{hh}) will be incorporated into $|{\Phi} _{\mathrm{G}}\rangle$ by fermionic statistical signs (see below). As the ground state after the duality transformation, $|{\Phi} _{\mathrm{G}}\rangle$ is presumably much smoother such that a perturbative treatment becomes possible.

Note that the phase string $\tau_C$ in (\ref{ps}) is closed-path-dependent as well as hole-spin-exchange-dependent. It thus resembles the mutual statistics. One may require $\hat{K}$ to keep track of the nonlocal phase shift in (\ref{ps}) as a generalized Berry's phase:
\begin{equation}
-i\oint_C d{\bf R}\cdot  \hat{K}^{\dagger}\partial_{\bf R}\hat{K}\rightarrow \pi N_h^{\downarrow}(\{C\}) ~,
\end{equation}
where ${\bf R}$ represents the mult-coordinates of the doped holes transversing the set of closed paths $\{C\}$. Such a $\hat{K}$ in the coordinate space can be written as
\begin{equation}
\hat{K}^{\dagger}\leftrightarrow\prod_{hd}\frac{z_{l_h}-z_{j_d}}{|z_{l_h}-z_{j_d}|}~,
\end{equation}
where $z$ is the complex coordinate with the subscripts $\{l_h\}$ and $\{j_d\}$ denoting the hole and down spin sites, respectively. In the second-quantization formalism, one has 
\begin{equation}
\hat{K}\equiv e^{i\hat{\Theta}}~ ,
\label{unitary2}
\end{equation}
with
\begin{equation}
\hat{\Theta} \equiv -\sum_i n_i^h \hat{\Omega}_{i} ~ .
\label{unitary3}
\end{equation}
Here, $n_i^h$ in $\hat{\Theta}$ defines the number operator of the hole created at site $i$ by the electron annihilation operator $\hat{c}$. Each doped hole introduces a nonlocal phase shift $\hat{\Omega}_{i}$, which is defined by
\begin{equation}
\hat{\Omega}_{i}\equiv \frac{1}{2} \sum_{l\neq i}\theta _{i}(l)\left( \sum_{\sigma }\sigma
n_{l\sigma }^{b}-1 \right)  \label{phif}
\end{equation}%
in which $\theta _{i}(l)=\mathrm{Im}\ln $ $(z_{i}-z_{l})$ and $n_{l\sigma }^{b}$ denotes the background spin number at site $l$. 

In this new world of Mottness, at the electron density close to the so-called half-filling with each lattice site occupied by one electron, the relevant degrees of freedom are collective ones. The charge carriers are at those sites where the electron numbers deviate from the half-filling; and the neutral spins are at half-filled sites. The doubly occupied sites are in a high-energy sector which only exist in virtual processes to mediate the so-called antiferromagnetic superexchange coupling between the nearest neighboring neutral spins in the large-$U$ limit. Generally speaking, the original single electrons do not directly exist as they are strongly coupled to each other as the Hilbert space is reduced by the no double occupancy constraint.  The mathematical question is how those low-energy degrees of freedom behave quantum mechanically; and the physical question is how the system responds to an external experimental probe. The goal is to search for a correct class of ground state wave functions when the conventional perturbative approach is totally invalidated. 

With the singular sign structure accurately embedded/regulated in $\hat{K}$ in the ground state $|\Psi _{\mathrm{G}}\rangle$, a mean-field-type treatment of the non-singular $|{\Phi} _{\mathrm{G}}\rangle $ is possible. At finite doping, it may be generally constructed in the following form \cite{Weng2011}
\begin{equation}
|{\Phi} _{\mathrm{G}}\rangle \equiv {\cal C} \exp \left(\sum_{ij}g_{ij}\hat{c}_{i\uparrow }\hat{c}_{j\downarrow } \right) |b\text{-RVB}
\rangle ~ . \label{frac1}
\end{equation}%
At half-filling, with $g_{ij}=0$, $|{\Phi} _{\mathrm{G}}\rangle $ recover the so-called bosonic RVB state $|b\text{-RVB} \rangle$, which was first proposed by Liang, Doucot, Anderson\cite{LDA} to accurately describe the antiferromagnetic state of the Heisenberg model. Note that at finite doping, the no double occupancy constraint in (\ref{frac1}) is naturally implemented because holes are created in pairs via $\hat{c}_{i\uparrow }\hat{c}_{j\downarrow } $ on the single-occupancy ``vacuum'' state $|b\text{-RVB} \rangle$. The hole pairing amplidue $g_{ij}$ is taken as a smooth variational parameter. 

At finite doping, the single-occupancy ``vacuum'' state $|b\text{-RVB} \rangle$ will be motified self-consistently by doping. In general, it may be written as $|b\text{-RVB}\rangle\equiv \hat{P}_s|\Phi _{b}\rangle $ with $\hat{P}_{s}$ enforcing the single-occupancy constraint $\sum_{\sigma} n_{i\sigma }^{b}=1$ on the bosonic RVB state
\begin{equation}
|\Phi _{b}\rangle \equiv \exp \left( \sum_{ij}W_{ij}b_{i\uparrow }^{\dagger
}b_{j\downarrow }^{\dagger }\right) |0\rangle _{b}~, \label{phirvb}
\end{equation}
with $n_{i\sigma }^{b}\equiv b_{i\sigma }^{\dagger }b_{i\sigma }$. Here the RVB pairing amplitude $W_{ij}$ is another variational parameter, which will decides the RVB pairing size depending on the doping concentration.

\section{Emergent world of partons}

By explicitly incorporating the non-integrable phase factor or the sign structure identified for the $t$-$J$ model, a new class of variational wave functions has been  constructed in (\ref{unitary}). In the following, we discuss the general features and consequences of such new type of ground states, which is clearly distinct from a conventional BCS wave function for superconductivity and a Fermi liquid state for normal state. 

\subsection{Fractionalization and parton description}

One may straightforwardly recast the ground state ansatz (\ref{unitary}) into a direct product state:
\begin{equation}
|\Psi _{\mathrm{G}}\rangle =\hat{P}\left( |\Phi _{h}\rangle \otimes |\Phi
_{a}\rangle \otimes |\Phi _{b}\rangle \right)  \label{gsansatz}
\end{equation}%
by decomposing the electron annihilation operator $c_{i\sigma }=\hat{P}\tilde{c}_{i\sigma }$
as follows\cite{Weng2011} 
\begin{equation}
\tilde{c}_{i\sigma }\equiv h_{i}^{\dagger }a_{i\bar{\sigma}}^{\dagger
}(-\sigma )^{i}e^{i\hat{\Omega}_{i}}\text{ }.  \label{decomp}
\end{equation}
The ground state (\ref{gsansatz}) is composed of three `fractionalized' subsystems. Here
\begin{equation}
|\Phi _{h}\rangle \equiv \sum_{\{l_{h}\}}\varphi
_{h}(l_{1},l_{2},...)h_{l_{1}}^{\dagger }h_{l_{2}}^{\dagger }...|0\rangle
_{h}\ ,  \label{bgs}
\end{equation}%
where the bosonic wavefunction $\varphi _{h}= const. $ defines
a Bose-condensed \textquoteleft holon\textquoteright \ state with a \emph{bosonic}\
creation operator $h_{l}^{\dagger }$ acting on a vacuum $|0\rangle _{h}$;
and
\begin{equation}
|\Phi _{a}\rangle \equiv \exp \left( \sum_{ij}{g}_{ij}(-1)^ia_{i\downarrow
}^{\dagger }a_{j\uparrow }^{\dagger }\right) |0\rangle _{a}~,  \label{phia-0}
\end{equation}%
which describes a \textquoteleft backflow spinon\textquoteright \ state with the `BCS'-like pairing
amplitude ${g}_{ij}$, where $a_{i\sigma }^{\dagger }$ denotes a \emph{fermionic} creation operator acting
on a vacuum $|0\rangle _{a}$; note that the bosonic RVB state $|\Phi _{b}\rangle $ is already given in Eq. (\ref{phirvb}). Finally, both (\ref{gsansatz}) and (\ref{decomp}) involve a projection operator, which is defined by
\begin{equation}
\hat{P}\equiv \hat{P}_{\mathrm{B}}\hat{P}_{s}\text{ ,}  \label{P}
\end{equation}%
in which $\hat{%
P}_{\mathrm{B}}$ will further enforce $n_{i\bar{\sigma}}^{a}=n_{i}^{h}n_{i\sigma }^{b}$,
such that each $a$-spinon always coincides with a holon as $\sum_{\sigma
}n_{i\bar{\sigma}}^{a}=n_{i}^{h}$  (here $n_{i\bar{\sigma}}^{a}\equiv a_{i\bar{\sigma}%
}^{\dagger }a_{i\bar{\sigma}}$ and $n_{i}^{h}\equiv h_{i}^{\dagger }h_{i}$
with $\bar{\sigma}\equiv -\sigma )$. By applying $\hat{P}$, the physical
Hilbert space is restored in (\ref{gsansatz}).

So we see that the original ground state ansatz (\ref{unitary}) is actually simplified as a direct product state of three subsystems in (\ref{gsansatz}). Here we do not see the trace of electrons directly anymore. Instead, they are all fractionalized into three types of pastons, the holons created by $h^{\dagger}$, the backflow fermionic spinons created by $a^{\dagger}$, and bosonic spinons created by $b^{\dagger}$. In particular, all of these partons are in the ODLRO states of their own in (\ref{bgs}), (\ref{phia-0}) and (\ref{phirvb}), respectively, where the holons are Bose condensed, $a$-spinons and $b$-spinons are in RVB paired states. 

It is well known in condensed matter physics that when a many-body quantum system exhibits an ODLRO by spontaneously breaking a global symmetry, it will become `rigid' against fluctuations attempting to destroy the ODLRO to restore the symmetry. For example, a crystal with a translational symmetry breaking naturally exhibits the rigidity in a conventional sense; a superconductivty ODLRO leads to the expulsion of external magnetic fields, etc. In other words, with the three parton subsystems exhibiting their own ODLROs, the conventional gauge fluctuations associated with the fractionalization (which would otherwise play a crital role to confine the partons should the decomposition is unphysical)  will get suppressed, which in turn justifies the present fractionalization scheme given in (\ref{decomp}). 

\subsection{Emergent gauge fields and the phase diagram}

It is important to note that the emergent gauge structure due to the non-integrable phase (sign) factor (\ref{ps}) is an intrinsic one, which cannot be `Higgsed' by the ODLROs in the parton subsystems mentioned above. Such an emergent gauge structure should be distinguished from the usual gauge fluctuations due to fractionalization, which are `Higgsed' by the ODRLOs to self-consistently support the fractionalization in (\ref{decomp}) as already pointed out in the Introduction.

Therefore, these partons are not simply described by some mean-field equations at finite doping. As a matter of fact, the fractionalization is precisely dictated by the emergent sign structure as discussed above. The quantum interference of these non-integrable phases from different paths will be characterized by a pair of emergent gauge fields, known as the mutual Chern-Simons gauge fields as discussed in Sec. II. Such a topological gauge force between the partons cannot be gauged away by the unitary transformation $\hat{K}$ and will play important roles in shaping the transformed ground state $|\Phi_G\rangle$ in (\ref{unitary}). 

In general, in the parton representation (\ref{gsansatz}), the emergent mutual Chern-Simons gauge fields will couple to the holons and spinons such that even though the ground state is approximately expressed in a direct product form, the mutual influences have been factored into each subsystem\cite{KQW2005,Ye2011}. In other words, the gauge force is essential in determining the ground states of the subsystems and their doping depdendences. In the following, we briefly outline some key features, while one is referred to Refs.\onlinecite{Weng2011,Ma2013} for more details. \\

\emph{Antiferromagnetic state, superconducting state, and the pseudogap physics}\\

At half-filling, all the electrons are reduced to neutral spins in the low-energy sector below the Mott gap \emph{without} the play of the fermionic
statistics. The ground state (\ref{unitary}) or (\ref{gsansatz}) is naturally turned into an antiferromagnetic state with a true magnetic
ODLRO in $|b\text{-RVB}\rangle$. $|b\text{-RVB}\rangle$ in this limit describes the ground state of the Heisenberg model very accurately as a variational state \cite{LDA}, which serves as the reference state for us to understand the finite-doped Mott insulator.  

At finite doping, the charge degree of freedom is introduced by doped holes in a background of neutral spins, and thus the particular fractionalization in Sec. IVA emerges due to the phase string like sign structure (\ref{ps}). A superconducting state arises once the localized doped holes in the dilute limit (cf. Sec. II) undergo a delocalization transition beyond some critical doping. 

The superconducting state of the doped Mott insulator is a natural ground state of pure electrons, without needing an extra
`gluon' like phonon in a BCS superconductor. In the latter, a Fermi liquid state is a natural ground state for purely electronic degrees of freedom, which sets in as a `normal state' once the Cooper pairing mediated by phonons is turned off. 

In this sense, the superconducting ground state (\ref{gsansatz}) is a stable infrared fixed point state, which also controls all the anomalous pseudogap
behavior at finite temperature. In other words, the basic correlations exhibited in high-temperature `normal state' regimes
are already encoded in the ground state, in a form of electron fractionalization. As shown in (\ref{gsansatz}), the fundamental degrees of freedom
are fractionalized into holons and two-component spinons in the ground state, which may be regarded as the `parent state'.

The so-called lower pseudogap phase (LPP) is formally very similar to the superconducting state in that the holons remain condensed, while both the $b$- and $a$-spinons are still RVB-paired. But with unpaired $b$-spinons thermally excited in $|\Phi _{b}\rangle $ at $T>0$, the superconducting phase coherence in the ground state (\ref{gsansatz}) is disordered in the LPP, because these thermally excited spinons automatically carry vortices, due to the novel sign structure as can be explicitly
seen in (\ref{unitary3}) and (\ref{phif})\cite{Ma2013}. Namely, the LPP state is a phase-disordered superconducting state, whose anomalous physical properties are thus
intimately tied to the non-BCS structure of the superconducting state itself.

Here the spin pseudogap behavior is manifested by $|\Phi _{b}\rangle $ \cite{Ma2013}. The bosonic RVB pairing of the $b$-spinons accurately depicts
the tendency for the local spins to develop antiferromagnetic correlations with decreasing temperature, which explains the high-temperature spin
pseudogap behavior over a wide range. 

On the other hand, the holon condensation would simply mean that the doped charges (holons) gain coherence. But due to the altered statistical sign
structure mentioned above, there generally exists a novel `mutual entanglement' between the holon and $b$-spinon subsystems, via the mutual
Chern-Simons gauge fields. Consequently, the spin excitations (the $b$-spinons) strongly affect the charge condensate by creating supercurrent vortices, i.e., the spinon-vortices. In the LPP, they disorder the superconducting phase coherence, resulting in a large non-Drude resistivity and strong Nernst effect as the
characteristics of non-Gaussian-like superconducting fluctuations in the LPP \cite{Ma2013}.

Self-consistently, the holon condensation in the LPP induces a spin gap $E_g$ in the spinon excitation spectrum and destroys the AFLRO. At a sufficiently low temperature, with the thermally excited spinon-vortices greatly reduced in number due to the spin gap, the confinement of them into vortex-antivortex pairs eventually becomes possible in the LPP, in a fashion of Kosterlitz-Thouless-type transition, which results in a true superconducting phase coherence below $T\leq T_c$ as controled by the spin gap $E_g$.

Another prediction for the ansatz (\ref{gsansatz}) is the existence of  a distinct pseudogap phase known as the LPP-II state \cite{Ma2013}. Such a state can be realized when the BCS-like pairing of the $a$-spinons in $|\Phi _{a}\rangle $ of  (\ref{phia-0}) is destroyed
(say, by strong magnetic fields at low $T$) \emph{before} the occurrence of phase disordering by thermally excited $b$-spinon-vortex excitations.
Correspondingly, the Cooper pairing amplitude vanishes to result in a non-superconducting normal state, at least in the magnetic vortex core region.

Note that the $a$-spinon in the superconducting phase and the LPP state is charge-neutral as well as gauge-neutral, immune from the mutual Chern-Simons gauge
force between the $b$-spinons and holons\cite{Ma2013}. It is gapped in the LPP such that its contributions to spin susceptibility and
specific heat vanish at low temperature. But in the LPP-II, the Fermi pockets of the fermionic $a$-spinons give rise to quantum oscillation, a Pauli
susceptibility and a linear-$T$ specific heat just like in a typical Fermi liquid. 

\section{Conclusion}

We have presented an example of emerging physics in a strongly correlated system of condensed matter. In the low-energy world of emergence, the underlying physical law has been `changed' from a conventional description of Bloch electrons with strong local repulsion to a novel one that is described by collective degrees of freedom, i.e., some `new' elementary particles, mutually interacting through an emergent gauge force. 

For a physicist living in this low-energy world without being able to get access to energies higher than the Mott gap,  he/she could not experimentally identify individual electrons as the basic constituents, and thus no Hubbard model could be deduced as the underlying physical law. Instead, the greatest triumph that he/she could possibly achieve is to `discover' the new elementary particles and a topological gauge structure as dictated by (\ref{ps}), which constitute the fundamental and complete physical law of this new universe.  This emergent world of gauge force is protected by the Mott gap, and is categorically different from the condensed matter universe that we know of at a `higher energy' level. Given the exactness of the mapping between the two worlds, it provides a perfect example to illustrate the idea of `more is different' in condensed matter physics.\cite{pwa1972,LP2000}

\section*{Acknowledgments}

This work was initiated during the workshop in celebration of C N Yang's 90th birthday and the 15th anniversary of Institute for Advanced Study at Tsinghua University (IASTU). We would like to thank the colleagues and students from IASTU for the earlier collaborations and discussions related to the present paper. The support from NSFC and MOST grants is also acknowledged.


\begin{thebibliography}{99}

\bibitem{Yang1974} C. N. Yang, Phys. Rev. Lett. \textbf{33}, 445 (1974); T. T. Wu and C. N. Yang, Phys. Rev. D \textbf{12}, 3845 (1975).
\bibitem{AGD1963} A. A. Abrikosov, L. P. Gor'kov, and E. Dzyaloshinskii, \emph{Method of Quantm Field Theory in Statistical Physics }, (Prentice-Hall, Englewood Cliffs, 1963). 
\bibitem{Pines1965} D. Pines and P. Nozieres, \emph{Theory of Quantum Liquids }, (Benjamin, New York 1965). 
\bibitem{pwa}  P. W. Anderson, Science \textbf{235}, 1196 (1987).
\bibitem{pwa1}  P. W. Anderson, \emph{The Theory of
Superconductivity in the High $T_c$ Cuprates}, (Princeton Univ. Press,
Princeton, 1997).
\bibitem{LNW2006} For a review, see, P. A. Lee, N. Nagaosa, and X. G. Wen,
Rev. Mod. Phys. \textbf{78}, 17 (2006).
\bibitem{Wang2012} C. Ye, P. Cai, R. Z. Yu, X. D. Zhou, W. Ruan, Q. Q. Liu, C. Q. Jin, and Y. Y. Wang, Nat. Commun. 4, 1365 (2013)


\bibitem{Weng1996} D. N. Sheng, Y. C. Chen, and Z. Y. Weng, Phys. Rev. Lett. 
\textbf{77}, 5102 (1996).

\bibitem{Weng1997} Z. Y. Weng, D. N. Sheng, Y.-Chen, and C. S. Ting, Phys.
Rev. B \textbf{55}, 3894 (1997).

\bibitem{WWZ2008} K. Wu, Z. Y. Weng, and J. Zaanen, Phys. Rev. B \textbf{77},
155102 (2008).

\bibitem{MW2010} J. W. Mei and Z. Y. Weng, Phys. Rev. B \textbf{81}, 014507
(2010).

\bibitem{Weng2007} Z. Y. Weng, Intl. J. Mod. Phys. B \textbf{21}, 773 (2007).

\bibitem{Zaanen2009} J. Zaanen and B. J. Overbosch, Phil. Trans. R. Soc. A 
\textbf{369}, 1599 (2011).

\bibitem{Weng2011r} Z. Y. Weng, Front. Phys. \textbf{6}, 370 (2011).

\bibitem{note1} So far, for simplicity, the whole discussion has been focused on the large-$U$ limit. But I wish to point out that essentially the same physics will persist over to a finite-$U$ so long as the splitting of the Bloch band into the lower and upper Hubbard bands by $U$ holds true, which will be presented elsewhere.\cite{Zhang2013}

\bibitem{Zhang2013} L. Zhang and Z. Y. Weng, \emph{unpublished}.  

\bibitem{KQW2005} S. P. Kou, X. L. Qi, and Z. Y. Weng,
Phys. Rev. B \textbf{71}, 235102 (2005).

\bibitem{Ye2011} P. Ye, C. S. Tian, X. L. Qi, and Z. Y. Weng, Phys. Rev.
Lett. \textbf{106}, 147002 (2011); Nucl. Phys. B \textbf{854}, 815 (2012).

\bibitem{Anderson1958} P. W. Anderson, Phys. Rev. \textbf{109}, 1492 (1958).

\bibitem{Weng2000} Z. Y. Weng, V. N. Muthukumar, D. N. Sheng, and C. S. Ting, Phys. Rev. B 
\textbf{63}, 075102 (2001).

\bibitem{White1992} S. R. White, Phys. Rev. Lett. \textbf{69}, 2863 (1992).

\bibitem{Zhu2012} Z. Zhu, H. C. Jiang, Y. Qi, C. S. Tian, and Z. Y. Weng, Scientific Reports, \textbf{3}, 2586 (2013).

\bibitem{Weng2011} Z. Y. Weng, New J. Phys, \textbf{13}, 103039 (2011).

\bibitem{LDA} S. Liang, B. Doucot, and P. W. Anderson, Phys. Rev. Lett. \textbf{61}, 365 (1988).

\bibitem{Ma2013} Y. Ma, P. Ye, and Z. Y. Weng, arXiv: 1311.3395.

\bibitem{pwa1972} P. W. Anderson, Science, \textbf{177}, 393 (1972). 

\bibitem{LP2000} R. B. Laughlin and D. Pines, PNAS, \textbf{97}, 28 (2000).

\end{thebibliography}
\end{document}